\documentclass{article}
\usepackage{spconf,amsmath,graphicx,amssymb}
\usepackage{xcolor}
\usepackage{booktabs}
\usepackage{multirow}
\usepackage{caption}
\usepackage{subcaption}
\usepackage{ownstyles}
\usepackage{enumitem}
\usepackage{mathtools}

\newif\ifcomments

\commentstrue

\ifcomments
\newcommand{\comments}[1]{#1}
\else
\newcommand{\comments}[1]{}
\fi

\newcommand*{\R}{\mathbb{R}}

\title{Improving Robustness to Adversarial Examples by Encouraging Discriminative Features
}
\name{Chirag Agarwal$^{\star}$ \qquad Anh Nguyen$^{\dagger}$ \qquad Dan Schonfeld$^{\star}$}
\address{$^{\star}$ University of Illinois at Chicago, Chicago, IL, USA 60607 \\
    $^{\dagger}$}\address{$^{\star}$ University of Illinois at Chicago, Chicago, IL, USA 60607 \\
    $^{\dagger}$ Auburn University, Auburn, AL, USA 36849}

\begin{document}
\maketitle
\begin{abstract}

Deep neural networks (DNNs) have achieved state-of-the-art results in various pattern recognition tasks. However, they perform poorly on out-of-distribution adversarial examples i.e. inputs that are specifically crafted by an adversary to cause DNNs to misbehave, questioning the security and reliability of applications.
In this paper, we hypothesize inter-class and intra-class feature variances to be one of the reasons behind the existence of adversarial examples. Additionally, learning low intra-class and high inter-class feature variance help classifiers learn decision boundaries that are more compact and leave less inter-class low-probability ``pockets'' in the feature space, i.e. less room for adversarial perturbations. We achieve this by imposing a center loss \cite{wen2016discriminative} in addition to the regular softmax cross-entropy loss while training a DNN classifier. Intuitively, the center loss encourages DNNs to simultaneously learn a center for the deep features of each class, and minimize the distances between the intra-class deep features and their corresponding class centers. Our results on state-of-the-art architectures tested on MNIST, CIFAR-10, and CIFAR-100 datasets confirm our hypothesis and highlight the importance of discriminative features in the existence of adversarial examples.

\end{abstract}

\section{Introduction}
\label{sec:intro}

In recent years, machine learning and in particular deep learning has impacted various fields, such as computer vision, natural language processing, and sentiment analysis. 
Regardless of their superior performances, these algorithms can be breached using adversarial inputs: perturbed inputs to force an algorithm to provide adversary-selected outputs \cite{szegedy2013intriguing}. 
Detailing all adversarial attacks in the literature is outside the scope of this paper. Hence, we will be providing a brief summary of different adversarial attacks, defenses and hypothesized reasons behind the existence of adversarial samples. 
Fast Gradient Sign Method (FGSM) was proposed as a fast method to generate adversarial samples by adding perturbation proportional to the sign of the cost functions gradient \cite{goodfellow6572explaining}. However, using iterative optimization-based attacks, such as Basic Iterative Method (BIM) \cite{kurakin2016adversarial}, Projected Gradient Descent (PGD) \cite{madry2017towards}, and Carlini-Wagner (CW) attack \cite{carlini2016towards}, have become standard practice
in evaluating defenses.
Black-box attacks (e.g. using numerical methods \cite{alcorn2019strike} or evolutionary algorithms \cite{su2017one,nguyen2015deep} to approximate the gradients) 
are more practical alternatives to white-box attacks as they do not assume that the attacker has access to the model weights or architectures for generating adversaries.
While many effective attack methods have been proposed~\cite{akhtar2018threat} e.g. single-pixel \cite{su2017one}, Local Search \cite{narodytska2016simple}, Houdini attack \cite{cisse2017houdini}, defending against adversarial examples (AXs) remains largely an open problem.


Recent approaches for defending against AXs include: (1) adversarial training \cite{goodfellow6572explaining}; (2) input transformation \cite{kabilan2018vectordefense}; (3) regularization \cite{rozsa2016towards}; and (4) detecting out-of-distribution inputs \cite{sabour2015adversarial}. 
Furthermore, there have been contrasting views on the existence of AXs. Linearity hypothesis, the flatness of decision boundaries, the large local curvature of the decision boundaries, or low flexibility of the networks are some viewpoints on the existence of AXs that do not perfectly align with each other \cite{akhtar2018threat}.

Inspired by k-Nearest Neighbor (kNN), classifiers that are trained with center loss \cite{wen2016discriminative} learn a center (feature vector) for each class. Intuitively, softmax cross-entropy loss attempts to maximize the inter-class distance and center loss attempts to minimize the distance between the intra-class features. We hypothesize that minimizing intra-class and maximizing inter-class feature distances would improve a classifier's robustness to AXs. In this paper, we show that DNN classifiers trained with a combination of center loss and softmax cross-entropy loss ($D_{SC}$) are substantially more robust to both white-box and black-box adversarial attacks compared to classifiers trained with softmax cross-entropy loss alone ($D_S$). The paper makes the following contributions:
\vspace*{-0.2em}
\begin{itemize}
\item Across MNIST, CIFAR-10, and CIFAR-100 datasets, our proposed models are more resistant to AXs generated by state-of-the-art white-box and black-box attacks including PGD, CW, and single-pixel attacks. 
\item The discriminative center loss when combined with the adversarial training \cite{goodfellow6572explaining} also improve the DNN robustness compared to vanilla adversarial training (Sec.~\ref{sec:adv_train}).
\end{itemize}
\section{A Feature Learning View on Robustness}
\label{sec:method}
Much of our discussion will revolve around how learning distinct features improves the adversarial robustness of a DNN. This study highlights how the approach can be used to achieve local robustness within the targeted classes. We provide the key details and notations that will be used in this section.

\vspace{0.5em}
\noindent\textbf{Notations.} Let $D(X) = Y$ represent a DNN used for a given classification task for any input image $X$, and $Y$ being the probability scores over the classes. The DNN maps an image to a probability distribution over the classes. We define the feature layer, the layer before the softmax layer, as $F$. So, the output probability score for the image $X$ is given by 
\begin{equation}
    D(X) = softmax(F(X))
\end{equation}The predicted class for the image $X$ is then given as $C(X) = arg max_{i} D(x_{i})$, where $D(x_{i})$ is the predicted class probability for class $i$. The adversarial examples, $X_{a}$, are samples that are adjacent to clean examples but the predicted label by the network is different, i.e. $C(X_{a}) \neq C(X)$. We now define the \textit{local robustness} for a DNN \cite{katz2017towards}.

\vspace{0.5em}
\noindent \textbf{Definition 1} A DNN $D$ is $\delta$ locally robust at point $x$ if 
\begin{equation}
\label{eq:local}
    \forall x_{a} \hspace{1.1em} ||x - x_{a}||_{2}^{2}\leq\delta\implies C(x) = C(x_{a})
\end{equation}

\begin{figure}[htb]
    \begin{subfigure}[b]{0.5\linewidth}
        \centering
        \centerline{\includegraphics[width=5.0cm]{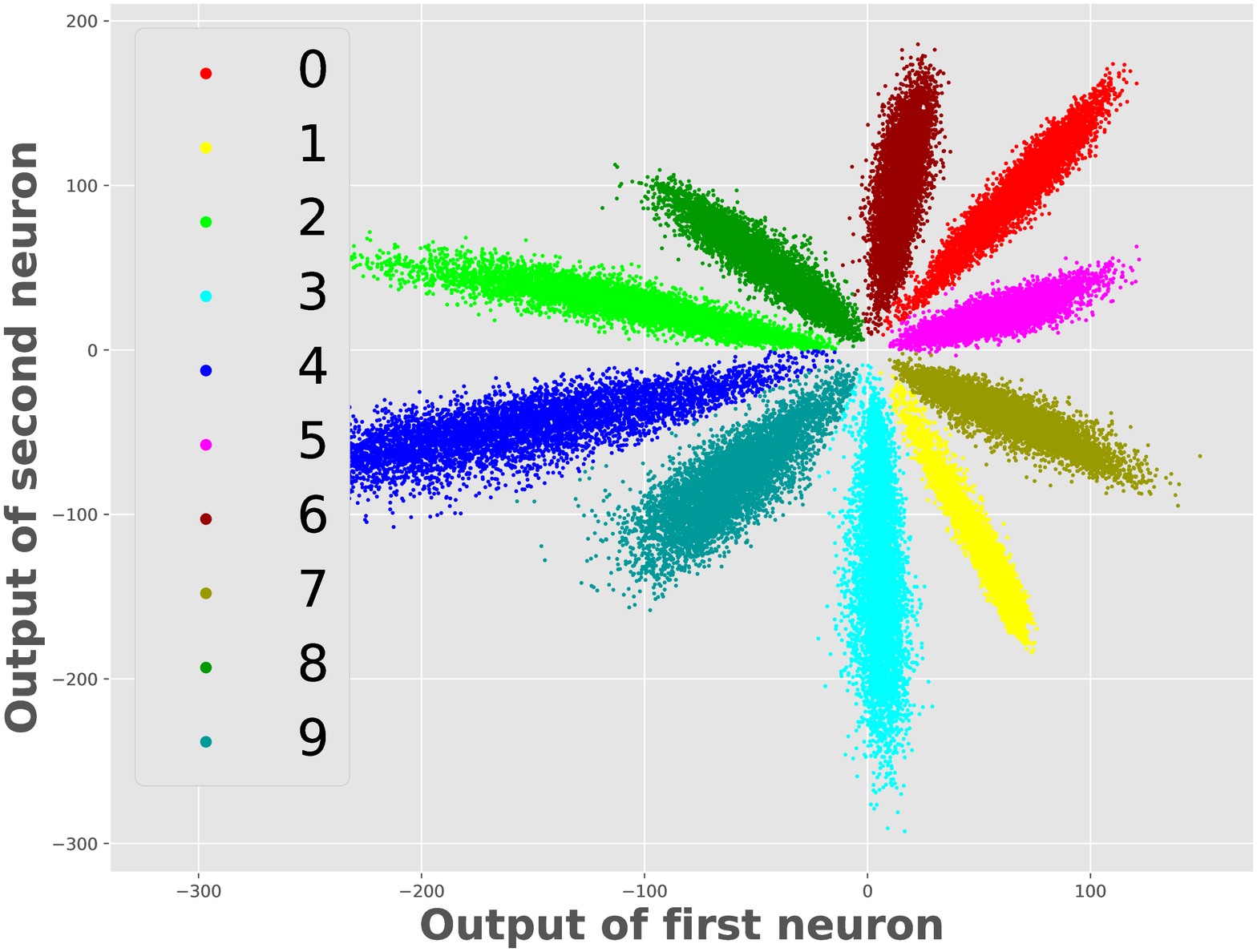}}
        \caption{$D_{S}$}
        \label{fig:fv_s}
    \end{subfigure}%
    \hfill
    \begin{subfigure}[b]{0.5\linewidth}
        \centering
        \centerline{\includegraphics[width=5.0cm]{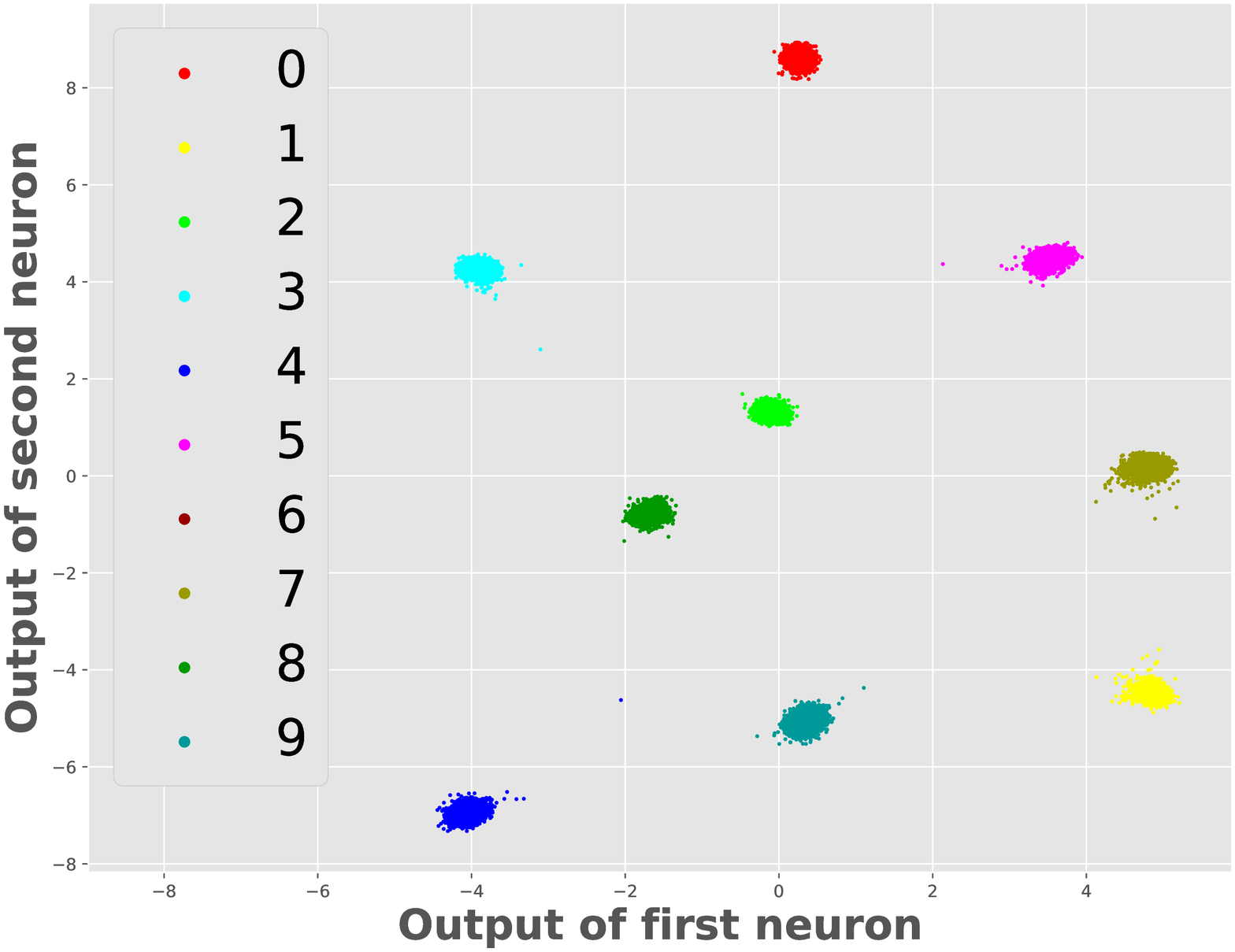}}
        \caption{$D_{SC}$}
        \label{fig:fv_sc}
    \end{subfigure}%
    \centering
\caption{Feature Visualization of the penultimate layer. $D_{SC}$ models results in higher inter-class and smaller intra-class variance.
}
\label{fig:fv}
\vspace*{-\baselineskip}
\end{figure}

\begin{figure}[t]
    \begin{subfigure}[b]{0.5\linewidth}
        \centering
        \centerline{\includegraphics[width=5.0cm]{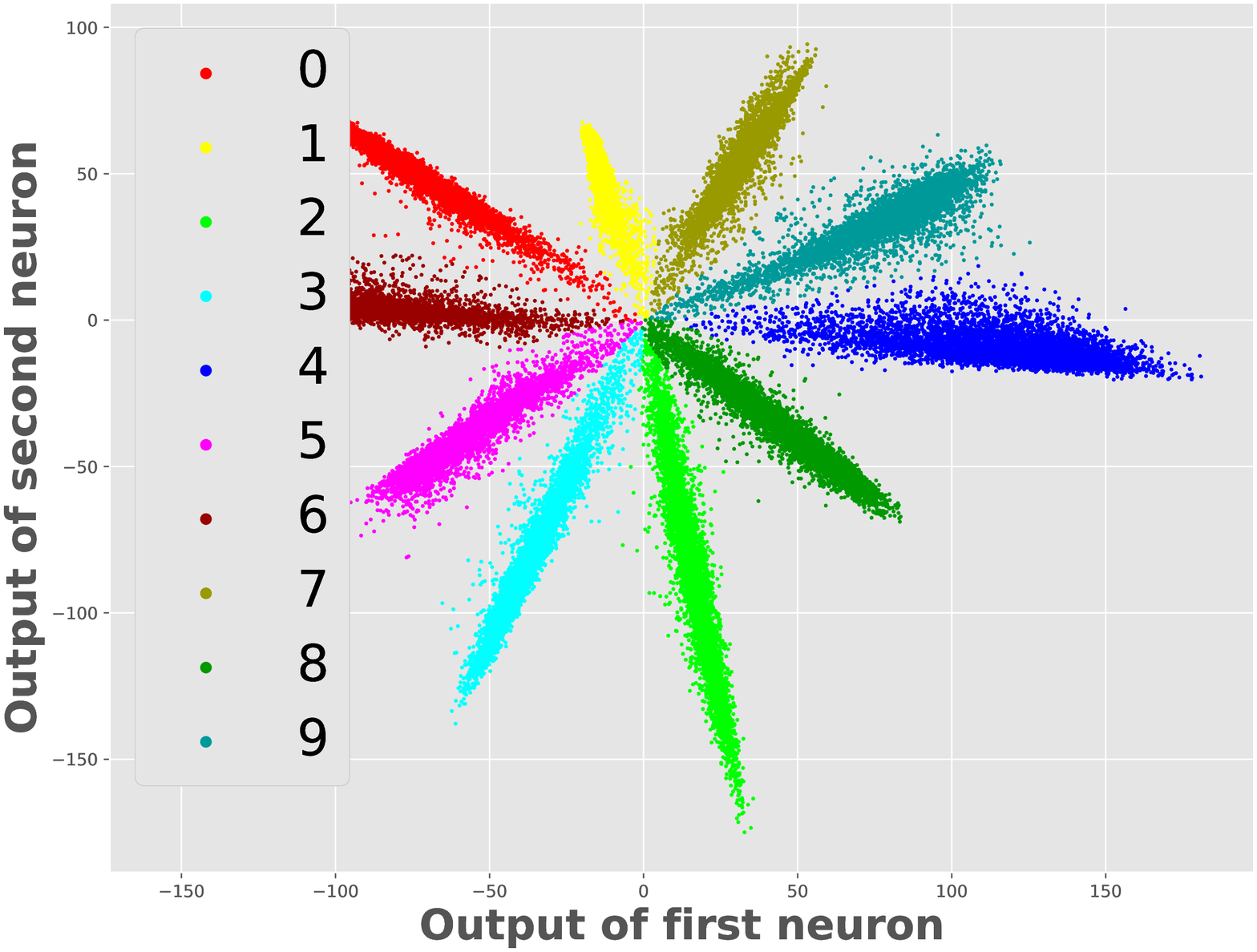}}
        \caption{$D_{AT\_S}$}
        \label{fig:fv_at_s}
    \end{subfigure}%
    \hfill
    \begin{subfigure}[b]{0.5\linewidth}
        \centering
        \centerline{\includegraphics[width=5.0cm]{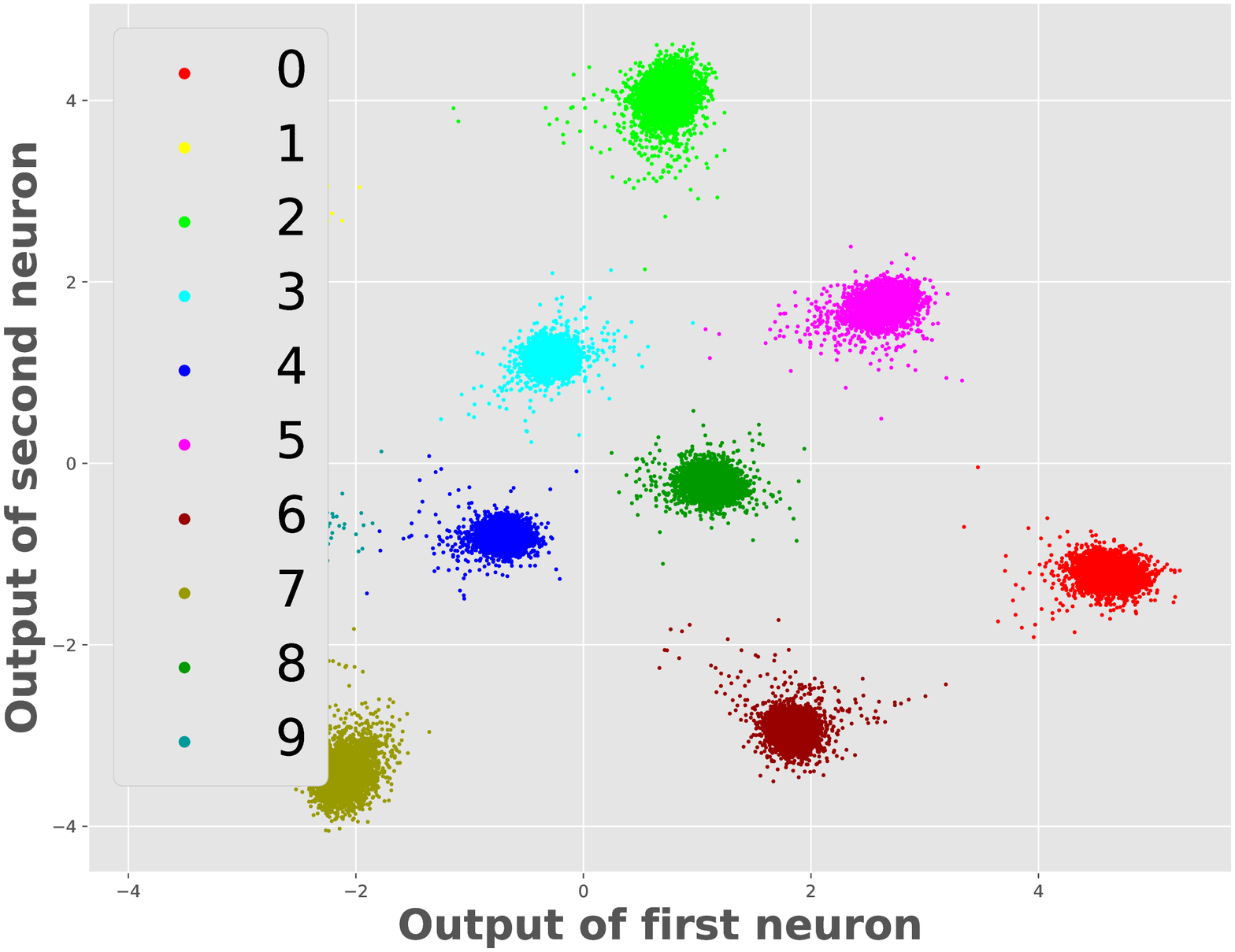}}
        \caption{$D_{AT\_SC}$}
        \label{fig:fv_at_sc}
    \end{subfigure}%
    \centering
\caption{Feature Visualization of the penultimate layer after adversarial training. The AXs cause the clean image features to intersect with each other and therefore causing the mis-classification
}
\label{fig:fv_at}
\vspace*{-\baselineskip}
\end{figure}

\noindent Eq. (\ref{eq:local}) inherently means that the DNN should assign the same labels to two input samples that are very close to each other, separated by $\delta$ in this case. Instinctively, this means that features learned from the same class should have smaller variance between them or in other words the features should have small intra-class variance. This gives us the motivation of using center-loss in addition to the softmax cross-entropy loss functions for classification.  Hence, all our DNN models are trained using the loss function defined in Eq. (\ref{eq:center}). For any input image $X_{k}$, the loss function is defined as:
\begin{equation}
\label{eq:center}
    L = -\sum_{k=1}^{m} \log \frac{e^{W^T_{y_{k}}F(X_k)}}{\sum_{j=1}^{n}e^{W^T_{y_{j}}F(X_k)}} + \lambda  \frac{1}{2} \sum_{k=1}^{m} || F(X_{k}) - C_{y_{k}}||^{2}_{2}
\end{equation}
where, $F(X_{k})$ denotes the feature layer output of the image $X_{k}$ with ground truth label $y_{k}$. $W_{j} \in \R^{d}$  denotes the $j$th column of the weights in $W \in \R^{d \times n}$ in the feature layer. The respective class centers are updated as the deep features change and are represented as $c_{y_{k}}$. The number of images and classes in the current dataset are denoted by $m$ and $n$ respectively. The value of $\lambda$ is taken as one for all our experiments \cite{wen2016discriminative}. Eq. (\ref{eq:center}) corresponds to the fact that features from the same class should have minimum distance between them and vice-versa. The local robustness can be now seen as the distance $\delta$ which represents the minimum distance to push a data point from one cluster to another. We explain this further using MNIST dataset \cite{lecun1998gradient} as our toy example. Similar to \cite{wen2016discriminative}, we train two versions of a LeNet network using ReLU activation with softmax and a combination of softmax and center loss separately. Hereafter, we denote a model trained with softmax loss as $D_{S}$ and the one with softmax and center loss together as $D_{SC}$. We train them using stochastic gradient descent (SGD) optimizer for 60k images and test on 10k testing samples. Additionally, we also used adversarial training to train both $D_{S}$ and $D_{SC}$ with AXs generated using FGSM attack with $\epsilon=0.3$. The adversarial trained models are represented by $D_{AT\_S}$ and $D_{AT\_SC}$ respectively. This choice of $\epsilon$ is the most common one in the context of adversarial examples and serves as a standard benchmark \cite{goodfellow6572explaining}. Fig. \ref{fig:fv} illustrates the feature learned by the feature layer (the penultimate layer) for the models $D_{S}$ (Fig. \ref{fig:fv_s}) and $D_{SC}$ (Fig. \ref{fig:fv_sc}). The results suggests that $D_{SC}$ learns more robust and discriminative features as compared to $D_{S}$. Subsequently, the minimum $\delta$ needed to generate AXs is smaller for Fig. \ref{fig:fv_s}. We validated this by attacking the trained models using FGSM attack algorithm and found that the performance of $D_{S}$ and $D_{SC}$ model falls from $96.95\%$ to $70.34\%$ and from $98.40\%$ to $78.73\%$ respectively. Clearly, $D_{SC}$ has a better adversarial performance than $D_{S}$. This point is further justified from Fig. \ref{fig:fv_s} where the high variance among intra-class features for the $D_{S}$ model is observed. 
\vspace*{-\baselineskip}
\vspace*{-0.25em}
One can increase the defensive power of an architecture by training the network with AXs that are generated on the fly. In Fig. \ref{fig:fv_at}, we visualize the features learned by $D_{AT\_S}$ and $D_{AT\_SC}$. On comparing Fig. \ref{fig:fv} and \ref{fig:fv_at}, it is interestingly seen that the inter-class distances become smaller due to the AXs. 
After adversarial training of the two models we tested the trained models against AXs generated using FGSM attack with $\epsilon=0.3$ and found that the $D_{AT\_SC}$ still performed better than $D_{AT\_S}$. This initial study (Table \ref{tab:toy}) solidifies our hypothesis that learning discriminative features increases the adversarial performance of a model.
\vspace*{-0.5em}
\begin{table}[htb]
\centering
\renewcommand{\arraystretch}{0.6}
\caption{Testing and Adversarial performance for the MNIST(LeNet model) using FGSM attack}
{\normalsize
\setlength{\tabcolsep}{9pt}
\begin{tabular}{|c|c|c|}
\hline
Model & Testing Accuracy & Adversarial Accuracy\\
\hline
\textbf{$D_{S}$} & 96.95\% & 70.34\% \\ \hline
\textbf{$D_{SC}$} & \textbf{98.40\%} & \textbf{78.93}\% \\ \hline
\textbf{$D_{AT\_S}$} & \textbf{98.94\%} & 96.08\%\\ \hline
\textbf{$D_{AT\_SC}$} & 98.73\% & \textbf{97.44}\% \\
\hline
\end{tabular}}
\vspace*{-\baselineskip}
\label{tab:toy}
\end{table}

\begin{table*}[]
\centering
\caption{Testing and Adversarial accuracy for different datasets, architectures and white-box attacks, such as FGSM, PGD and CW attack}
\label{tab:white}
{\small
\setlength{\tabcolsep}{1pt}
\resizebox{1.0\textwidth}{!}{%
\renewcommand{\arraystretch}{1.3}
\begin{tabular}{|c|c|c|c|c|c|c|c|c|c|c|c|c|}
\hline
\multirow{3}{*}{{Datasets}} & \multirow{3}{*}{{Model}} & \multicolumn{2}{c|}{\multirow{2}{*}{{\begin{tabular}[c]{@{}c@{}}Testing Accuracy\end{tabular}}}} & \multicolumn{9}{c|}{{\begin{tabular}[c]{@{}c@{}}Adversarial Accuracy\end{tabular}}} \\ \cline{5-13} 
 &  & \multicolumn{2}{c|}{} & \multicolumn{3}{c|}{{FGSM Attack}} & \multicolumn{3}{c|}{{PGD Attack}} & \multicolumn{3}{c|}{{CW Attack}} \\ \cline{3-13} 
 &  & \textbf{$D_{S}$} & \textbf{$D_{SC}$} & \textbf{$D_{S}$} & \textbf{$D_{SC}$} & $p$\textit{-val} & \textbf{$D_{S}$} & \textbf{$D_{SC}$} & $p$\textit{-val} & \textbf{$D_{S}$} & \textbf{$D_{SC}$} & $p$\textit{-val}  \\ \hline
{MNIST} & {(a) MLP} & $97.91 \pm 0.1\%$ & $97.97 \pm 0.2\%$ & $48.23 \pm 5.4\%$ & $68.68 \pm 3.6\%$ & \textless 0.001 & $0.01 \pm 0.0\%$ & $0.08 \pm 0.1\%$ & 0.002 & $0.03 \pm 0.0\%$ & $4.82 \pm 0.5\%$ & \textless 0.001 \\ \hline
\multirow{3}{*}{{CIFAR-10}} & {(b) VGG19} & $93.42 \pm 0.3\%$ & $92.36 \pm 0.2\%$ & $71.91 \pm 1.1\%$ & $69.56 \pm 2.3\%$ & 0.054 & $8.47 \pm 0.2\%$ & $27.61 \pm 1.4\%$ & \textless 0.001 & $8.98 \pm 2.4\%$ & $47.06 \pm 3.1\%$ & \textless 0.001 \\ \cline{2-13} 
 & {(c) ResNet18} & $94.97 \pm 0.3\%$ & $94.52 \pm 0.2\%$ & $68.64 \pm 2.2\%$ & $74.29 \pm 1.4\%$ & \textless 0.001 & $0.81 \pm 0.3\%$ & $29.26 \pm 1.0\%$ & \textless 0.001 & $ 0.00 \pm 0.0\%$ & $59.57 \pm 7.5\%$ & \textless 0.001\\ \cline{2-13} 
 & {(d) DenseNet40} & $94.37 \pm 0.2\%$ & $93.91 \pm 0.1\%$ & $55.44 \pm 0.8\%$ & $69.92 \pm 2.8\%$ & \textless 0.001 & $0.29 \pm 0.5\%$ & $23.91 \pm 3.5\%$ & \textless 0.001 & $0.00 \pm 0.1\%$ & $0.18 \pm 0.1\%$ & 0.006\\ \hline
\multirow{3}{*}{{CIFAR-100}} & {(e) VGG19} & $73.33 \pm 0.2\%$ & $69.10 \pm 1.5\%$ & $40.26 \pm 0.5\%$ & $38.63 \pm 3.2\%$ & 0.272 & $2.93 \pm 1.3\%$ & $9.04 \pm 2.0\%$ & \textless 0.001 & $8.45 \pm 1.7\%$ & $23.07 \pm 2.5\%$ & \textless 0.001\\ \cline{2-13} 
 & {(f) ResNet18} & $75.13 \pm 0.8\%$ & $76.91 \pm 0.2\%$ & $38.46 \pm 0.7\%$ & $43.43 \pm 1.2\%$ & \textless 0.001 & $0.32 \pm 0.2\%$ & $5.16 \pm 1.3\%$ & \textless 0.001 & $0.0 \pm 0.0\%$ & $17.49 \pm 1.5\%$ & \textless 0.001\\ \cline{2-13} 
 & {(g) DenseNet40} & $75.85 \pm 0.3\%$ & $73.78 \pm 0.3\%$ & $26.36 \pm 0.5\%$ & $32.08 \pm 0.4\%$ & \textless 0.001 & $0.07 \pm 0.0\%$ & $1.89 \pm 0.4\%$ & \textless 0.001 & $0.00 \pm 0.0\%$ & $0.75 \pm 0.3\%$ & 0.002\\ \hline
\end{tabular}%
}}
\vspace*{-\baselineskip}
\end{table*}
\section{Experiments and Results}
\label{sec:exp}
Building upon the results from Sec.~\ref{sec:method}, we now present extensive results that encompasses state-of-the-art white-box attacks, black-box attacks, and adversarial training method. Comprehensive experiments using widely used sequential and skip-connected architectures, along with popular classification datasets, such as MNIST, CIFAR-10, and CIFAR-100, are used to validate our hypothesis.
\subsection{Datasets and Networks}
\label{subsec:data}
\noindent\textbf{MNIST:} The experiments on toy example (Sec.~\ref{sec:method}) and other adversarial attacks were performed on the standard MNIST dataset of handwritten digits. Each sample is a $28\times28$ binary image. The complete dataset comprises of 70k images, divided into 60k training and 10k testing images. For pre-processing, we normalize the data using mean and standard deviation. For MNIST, we use a 3-layer ReLU network having 200 nodes in each hidden layer, similar to \cite{papernot2017practical}, as a standard MNIST architecture. The network was trained using SGD for 30 epochs. The initial learning rate is set to 0.1, and is divided by 10 at 50\% and 75\% of the total number of training epochs. No data augmentation was used for MNIST.\\

\noindent\textbf{CIFAR:} CIFAR dataset is a collection of two secondary datasets, CIFAR-10 and CIFAR-100, consisting $32\times32$ RGB images drawn from 10 and 100 classes respectively. Each of them contains 60k images split into 50k training images, and 10k testing images. We adopt a standard data augmentation scheme, random crop and flip.
CIFAR-10 and CIFAR-100 architectures are trained for 300 epochs respectively using the same training scheme as MNIST. For our experiments, VGG-19 \cite{simonyan2014very}, ResNet-18 \cite{he2015deep} and DenseNet-40-without-bottleneck \cite{huang2017densely} models are used. 
\subsection{White-box attacks}
\label{sec:white}
This type of adversarial attacks assume complete knowledge of the targeted model, including its parameters, architecture, training method, and in some cases its training data as well. Adversarial strength is defined by the certainty that the attacking algorithm would generate AXs which will be misclassified by the model. PGD and CW attacks are considered to be white-box attacks with high adversarial strengths. FGSM, on the other hand, is a quick way to generate low strength adversaries. In Table \ref{tab:white}, we summarize the performance of various architectures on these white-box attacks. In order to statistically validate, we train five different models for all architectures and perform two sample t-test on the adversarial accuracies between $D_S$ and $D_{SC}$. The $p$\textit{-val} in the table highlights the statistical significance of the results. All the architectures were trained based on clean examples. Surprisingly in the MNIST results, we see a huge difference in the adversarial performance between $D_{S}$ and $D_{SC}$ for AXs generated from FGSM with an increase of $\approx20\%$. We do not observe a significant difference for the PGD and CW attacks as they are very strong attacks and MNIST is a relatively easier dataset to fool. Despite the testing accuracy of $D_{SC}$ being lower than $D_{S}$, we observe that the adversarial performance of the former is significantly higher. In general, we get an increase in adversarial performance throughout all the attacks and architectures. Notably, for very strong attacks like PGD and CW, we observe significant increase in adversarial performance for $D_{SC}$ models in the CIFAR dataset results. For VGG-19 results of PGD and CW attack (Table \ref{tab:white} b--e), we see an increase as high as $\approx\textbf{487\%}$. A similar trend is seen in the case of CIFAR-10 ResNet-18 architecture results where an increase up to $\approx\textbf{500\%}$ and in some cases a change from $\textbf{0\%}$ to $\approx\textbf{60\%}$ (Table \ref{tab:white} c) is observed. Moreover, the $p$\textit{-val} $(\textless 0.05)$ in most of our comparisons statistically proves that the difference in their performance is significant. One should note that this increase in performance is achieved without using any AXs for training.
\begin{table}[]
\caption{Testing and Adversarial accuracy for different datasets, architectures using single-pixel, black-box attack.
}
\label{tab:black}
\centering
{\small
\setlength{\tabcolsep}{1pt}
\resizebox{0.5\textwidth}{!}{%
\renewcommand{\arraystretch}{1.6}
\begin{tabular}{|c|c|c|c|c|c|c|}
\hline
\multirow{3}{*}{{Datasets}} & \multirow{3}{*}{{Network}} & \multicolumn{2}{c|}{\multirow{2}{*}{{\begin{tabular}[c]{@{}c@{}}Testing Accuracy\end{tabular}}}} & \multicolumn{3}{c|}{\multirow{2}{*}{{\begin{tabular}[c]{@{}c@{}}Adversarial Accuracy\end{tabular}}}} \\ 
 &  & \multicolumn{2}{c|}{} & \multicolumn{3}{c|}{} \\ \cline{3-7} 
 &  & \textbf{$D_{S}$} & \textbf{$D_{SC}$} & \textbf{$D_{S}$} & \textbf{$D_{SC}$} & $p$\textit{-val} \\ \hline
{MNIST} & {(a) MLP} & $97.91 \pm 0.1\%$ & $97.97 \pm 0.2\%$ & $95.32 \pm 1.5\%$ & $94.69\pm0.9\%$ & 0.403 \\ \hline
\multirow{3}{*}{{CIFAR-10}} & {(b) VGG19} & $93.42 \pm 0.3\%$ & $92.36\pm0.2\%$ & $77.41\pm0.5\%$ & $78.50\pm0.7\%$ & 0.013\\ \cline{2-7} 
 & {(c) ResNet18} & $94.97 \pm 0.3\%$ & $94.52 \pm 0.2\%$ & $84.16\pm0.1\%$ & $84.68\pm0.5\%$ & 0.051\\ \cline{2-7} 
 & {(d) DenseNet40} & $94.37 \pm 0.2\%$ & $93.91 \pm 0.1\%$ & $81.16\pm0.7\%$ & $81.16\pm0.9\%$ & 1.0\\ \hline
\multirow{3}{*}{{CIFAR-100}} & {(e) VGG19} & $73.33 \pm 0.2\%$ & $69.10 \pm 1.5\%$ & $43.02\pm0.5\%$ & $45.01\pm0.3\%$ & \textless 0.001 \\ \cline{2-7} 
 & {(f) ResNet18} & $75.13 \pm 0.8\%$ & $76.91 \pm 0.2\%$ & $57.97\pm1.5\%$ & $56.18\pm0.9\%$ & 0.036\\ \cline{2-7} 
 & {(g) DenseNet40} & $75.85 \pm 0.3\%$ & $73.78 \pm 0.3\%$ & $57.13\pm3.1\%$ & $51.38\pm1.1\%$ & 0.005 \\ \hline
\end{tabular}}%
}\vspace*{-\baselineskip}
\end{table}
\subsection{Black-box attacks}
\label{subsec:black}
Black-box attacks generate AXs, during testing, for a targeted model without prior knowledge of model parameters. Some versions of black-box assume having some knowledge of the model architecture or the training process but in no case do they know about the model weights. The black-box attack performance is evaluated using the same trained models used in Sec.~\ref{sec:white}. In Table \ref{tab:black}, we see a notable difference in the adversarial performance for black-box attack. $D_{SC}$ models force the training procedure to learn more discriminative features. This inherently introduces a trade-off between the model's testing and adversarial accuracy. It is clearly seen, from Table \ref{tab:black} (a, d), that the difference in the adversarial performance for the MNIST and CIFAR-10 cases are non-significant. Moreover, for a tougher dataset like CIFAR-100 (Table \ref{tab:black} e--g) we observe the adversarial performance to be a by-product of the testing accuracy of the architectures on clean examples. The significant drop in $D_{SC}$ adversarial performance for CIFAR-100-DenseNet-40 is clearly observed as its testing accuracy was lower than the $D_{S}$ models. We believe regularizing the $D_{SC}$ models, for increasing testing accuracy on clean examples, would further improve their black-box adversarial performance.\vspace{-0.5em}
\subsection{Adversarial Training}
\label{sec:adv_train}
As shown in Sec.~\ref{sec:method}, adversarial training drastically improves the adversarial accuracy of any model. The adversarial training process is a brute-force approach and is very slow because we generate AXs on the fly. Hence, we choose the best models for each architecture, from Table \ref{tab:white}, and use it for our adversarial training experiments. Table \ref{tab:adv_train} illustrates the testing and adversarial performance of the previously chosen models (from  Sec.~\ref{subsec:data}) when we train them with both clean and adversarial samples, i.e. adversarial training. For MNIST, we train the MLP model for 30 epochs with the adversarial samples introduced on the fly after a delay of 10 epochs. The delay is important as we make sure that the network first achieves a decent performance with clean examples and in our case they achieve $90\%$ of the expected accuracy. We chose BIM attack for generating AXs during the training because it generates higher strength adversaries as compared to FGSM and is faster than CW attack \cite{akhtar2018threat}. Interestingly, MNIST, being a relatively easier dataset, is not affected due to the adversarial training and hence, the adversarial performance of $D_{AT\_S}$ and $D_{AT\_SC}$ are close. For CIFAR datasets, the models were trained for 300 epochs and the AXs were introduced after a delay of 150 epochs. The training using AXs affects the testing accuracy and hence we see notable difference in the testing accuracy between Table \ref{tab:white} and  \ref{tab:adv_train}. After adversarial training, the models were tested on new AXs generated from BIM attack and we see a significant difference in their subsequent adversarial performances. Notably, we see a significant increase in the adversarial accuracy throughout Table \ref{tab:adv_train}. Even adversarial training does not bridge the adversarial performance gap between $D_{S}$ and $D_{SC}$ models in Table \ref{tab:white} as the latter learns more discriminative features from the AXs too.
\begin{table}[hbt!]
\centering
\caption{Testing and Adversarial accuracy for different datasets and architectures after adversarial training using BIM white-box attack}
\label{tab:adv_train}
{\small
\setlength{\tabcolsep}{2pt}
\renewcommand{\arraystretch}{0.9}
\begin{tabular}{|c|l|c|c|c|c|}
\hline
\multirow{2}{*}{{Datasets}} & \multirow{2}{*}{{Network}} & \multicolumn{2}{c|}{{\begin{tabular}[c]{@{}c@{}}Testing Accuracy\end{tabular}}} & \multicolumn{2}{c|}{{\begin{tabular}[c]{@{}c@{}}Adversarial Accuracy\end{tabular}}} \\ \cline{3-6} 
 &  & \textbf{$D_{AT\_S}$} & \textbf{$D_{AT\_SC}$} & \textbf{$D_{AT\_S}$} & \textbf{$D_{AT\_SC}$} \\ \hline
{MNIST} & {(a) MLP} & 98.01\% & \textbf{98.69\%} & \textbf{94.33\%} & 94.22\% \\ \hline
\multirow{3}{*}{{CIFAR-10}} & {(b) VGG19} & \textbf{87.26\%} & 85.89\% & 61.39\% & \textbf{65.78\%} \\ \cline{2-6} 
 & {(c) ResNet18} & \textbf{89.30\%} & 88.43\% & 60.45\% & \textbf{66.10\%} \\ \cline{2-6} 
 & {(d) DenseNet40} & \textbf{88.19\%} & 86.02\% & 66.34\% & \textbf{72.59\%} \\ \hline
\multirow{3}{*}{{CIFAR-100}} & {(e) VGG19} & \textbf{70.10\%} & 69.89\% & 44.00\% & \textbf{47.85\%} \\ \cline{2-6} 
 & {(f) ResNet18} & 70.18\% & \textbf{71.25\%} & 47.68\% & \textbf{49.28\%} \\ \cline{2-6} 
 & {(g) DenseNet40} & \textbf{69.19\%} & 68.30\% & \textbf{44.48\%} & 35.86\% \\ \hline
\end{tabular}}
\vspace*{-\baselineskip}
\end{table}
\vspace*{-0.1em}
\section{Conclusion}
\label{sec:conclusion}
In this work, we hypothesized that one of the possible reasons behind the poor adversarial performance of the state-of-the-art deep learning architectures are their lack of learning discriminative deep features. We bridged that gap by using a combination of softmax and center loss function and then
performed a comprehensive set of experiments to successfully substantiate the effectiveness of our hypothesis. We plan to further investigate in this direction to propose defense mechanism for deep learning models.
\clearpage

%

\bibliographystyle{IEEEbib}
\bibliography{strings,refs}

\end{document}